\documentclass[3p]{elsarticle}%
\usepackage{amsmath}
\usepackage{amsfonts}
\usepackage{url}
\usepackage{optprog}
\usepackage{graphicx}
\usepackage{algorithm}
\usepackage{algpseudocode}
\usepackage{amsthm}

\usepackage[normalem]{ulem}

\newcommand\ignore[1]{}

\begin{document}
\title{Online Portfolio Selection: Uniform Dirichlet vs. Factor Dirichlet Weights}
\author[1]{Purushottam Parthasarathy\corref{cor}
}
\ead{194192001@iitb.ac.in}

\author[1,2]{Avinash Bhardwaj
}
\ead{abhardwaj@iitb.ac.in}
\author[1]{Manjesh K. Hanawal
}
\ead{mhanawal@iitb.ac.in}
\affiliation[1]{organization = {Department of Industrial Engineering and Operations Research},
addressline={IIT Bombay},
city={Mumbai, Maharashtra},
postcode={400076},
country={India}
}
\affiliation[2]{organization = {Department of Mechanical Engineering},
addressline={IIT Bombay},
city={Mumbai, Maharashtra},
postcode={400076},
country={India}
}
\cortext[cor]{Corresponding author}


\begin{abstract}

We revisit the problem of online portfolio allocation first introduced by Cover. We propose factor weighing of the dirichlet sampling to construct universal portfolios that out-perform those using uniform dirichlet sampling. When the returns follow a factor model, we establish a lower bound on the average portfolio growth rate. We analytically establish that the wealth generated by the factor weighted dirichlet sampled portfolios dominate the wealth generated by the uniformly sampled Dirichlet portfolios. We corroborate our analytical results with empirical studies on equity markets that are known to be driven by factors.  
\end{abstract}
\begin{keyword}
universal portfolio \sep portfolio optimization \sep online portfolio selection \sep dirichlet portfolio 
\end{keyword}
\maketitle


\section{Introduction}
Portfolio selection is an important and well studied problem across the disciplines of finance, economics, operations research and computer science. While there are many flavors to approaching this problem, two major streams of literature exist:
\begin{itemize}
    \item Single or Multi-period optimization: This body of work focuses on what can be done optimally from a mean variance trade-off perspective, in a single or multi-period context.
    \item Online portfolio selection: The literature here focuses on arriving at the best average growth rate for the portfolio over the long run.
\end{itemize}

In this work we explore the state of the art for online portfolio selection and propose a method for asset allocation in factor driven markets that provides an effective alternative to some of the complex techniques that have been proposed in the literature.

\subsection{Literature review}
Cover\cite{Cover91} shows that one can come up with an adapted strategy for online portfolio selection that asymptotically outperforms the best stock in the portfolio. The portfolio that achieves this is the performance weighted average of all the possible portfolios in the n-simplex, where performance is defined as the cumulative return till date. In fact, Cover's\cite{Cover91} central result is that the exponential growth rate of a `non-anticipating' fully adapted portfolio with no look-ahead bias (one that is proposed) asymptotically approaches the growth rate of a constant weight re-balanced portfolio with the benefit of hindsight. 
The update in each step tilts towards the portfolios that have performed well so far and continuously tracks a weighted average portfolio. Note that as Cover\cite{Cover91} restricts the benchmark portfolio to constant proportional weight re-balanced portfolios, it becomes clear that although \cite{Cover91}'s choice of a benchmark is reasonable within the context of what is presented, without this restriction of constant weight re-balancing a portfolio with the advantage of hindsight can achieve significantly higher returns\cite{singer1997switching}.
In subsequent work done as an extension to Cover\cite{Cover91}, Cover and Ordentlich\cite{cover1996universal} find a tight error bound on the maximum deviation between the best constant re-balanced portfolio in hindsight and the one they arrive at using side information without the benefit of hindsight. Side information is modelled in a discrete fashion here, selected from a set of k possible values. 
In related work \cite{helmbold1998line} show that exponential multiplicative updates yield almost the same performance as the universal portfolio and yet have less memory requirements. In \cite{helmbold1998line} the authors use a utility function that serves to trade-off a performance measure and a distance measure. The performance is based on $x_t$, the current relative price vector and the distance measure is chosen to be the relative entropy between the new weight vector and the current one. An approximation of the optimization objective function along with a constraint that the portfolio is fully invested leads to a recursive update algorithm that uses an exponential update to the weight vector. In various empirical tests, the algorithm achieves better out of sample performance than Cover's\cite{Cover91} portfolio and some other benchmarks.

As a sharp contrast to the universal portfolio type idea where best assets are assumed to perform well in the future, some approaches use a contrarian approach. The authors in \cite{AntiCor2003} show that a forecast step that uses mean-reversion can be used to get a performant portfolio. The authors calculate a transfer(i $\to$ j) which depends upon a few factors that mimic mean reversion and use lead-lag relationships.
The authors show that this heuristic does well empirically.

There is also a set of work that has been done that explores meta-algorithms (MAs). The authors in \cite{das2011meta} use daily S\&P500 data for the past 21 years to show that MAs outperform existing portfolio selection algorithms by several orders of magnitude.
As a good summary of available work, Li et al. \cite{li2014online} have written a comprehensive review paper that describes many of the major techniques in this area.

\section{Constant re-balanced portfolios in factor driven markets}
\subsection{Constant re-balanced portfolios for a single factor market }
In this section we find an analytical expression for the growth rate of a constant re-balanced portfolio in a market that is driven by a single factor. \ignore{This derivation serves to setup some of the preliminaries for results we show later.}We assume throughout that there are m assets, n periods of trading. A portfolio $b=(b_1,b_2,b_3...b_m), b_i \geq 0, \Sigma b_i = 1$ is the allocation of wealth to assets. The symbol t stands for transpose. $X_i$ is the $i^{th}$ relative price vector. A relative price is defined as the ratio of the price in the next time period to the current one.
The long term average growth rate of a portfolio is given as:

\begin{equation*}\label{eq:growth}
\frac{1}{n} \log (\prod_{i=1}^{n} b^t X_{i})
\end{equation*}

Consider a single factor model as:

\begin{equation*}\label{eq:factormodel}
\begin{bmatrix}
X_1^{i}\\
X_2^{i}\\
.\\
.\\
X_m^{i}\\
\end{bmatrix} = \begin{bmatrix}
\beta_{1} \; R_{m}^i\\
\beta_{2}\; R_{m}^i\\
.\\
.\\
\beta_{m}\; R_{m}^i\\
\end{bmatrix} =  R_{m}^i \; \boldsymbol{\beta} 
\end{equation*}
where $R_m^i$ is a scalar that denotes single-factor (market) returns for the $i^{th}$ period , and $\beta_{k}$ is a quantity that measures the factor exposure of the $k^{th}$ asset to the factor return. Using the above expression into the definition of the growth rate we have:

\begin{equation*}\label{eq:growth2}
G = \frac{1}{n} \log (\prod_{i=1}^{n} b^t R_{m}^i \boldsymbol{\beta}    )
\end{equation*}

\begin{equation*}\label{eq:growth3}
G = \frac{1}{n} \log ( \prod_{i=1}^{n} R_{m}^i \prod_{i=1}^{n} b^t  \boldsymbol{\beta}    )
\end{equation*}

\begin{equation*}\label{eq:growth4}
G = \frac{1}{n} \left[ \log ( \prod_{i=1}^{n} R_{m}^i ) + n\;  \log  b^t  \boldsymbol{\beta} \right]
\end{equation*}

\begin{equation*}\label{eq:growth5}
G = \log  b^t  \boldsymbol{\beta} + \frac{1}{n} \left[ \log ( \prod_{i=1}^{n} R_{m}^i ) \right] = \log  b^t  \boldsymbol{\beta} + G_{m}
\end{equation*}

which shows us that the excess average growth rate over the market growth rate $G_m$ is given by the first term in the above expression. The maximum growth rate is achieved at the vertex of the simplex which corresponds to the highest $\boldsymbol{\beta}$ asset, which is equivalent to holding investment at the asset with the highest $\boldsymbol{\beta}$.

\subsection{Growth Rate of multi-factor portfolios}
In this section we derive the relationship between the growth of a constant re-balanced portfolio and the growth of the underlying factors when the portfolio constituents can be assumed to follow a log-factor model.
A stock market as modelled in Cover\cite{cover1999elements} is represented as a vector of price-relatives that are assumed to follow a distribution $F(x)$. Define $X_i$ as the one-period price-relative for asset $i$, which is equal to $p^{end}_i/p^{start}_i$ where $p_i$'s are measured at the start and end of the time period (single period model). The wealth that is generated by engaging $b$ over the single period price-relatives is $S = b^tX$. As the investment model involves re-investing of the capital every period, we are interested in the cumulative product of these single period gains and thereby characterize the expected logarithm of this gain\cite{cover1999elements} as: 

\begin{equation}\label{eq:gain1}
W(b, F) = \int \log b^t x \; dF(x) = E\left[ \log b^tX \right]
\end{equation}

The optimal $b^*$ that maximizes (\ref{eq:gain1}) is also the doubling rate of the portfolio as shown in \cite{cover1999elements}.
We now consider a 2-factor log returns factor model of the market where known factors drive the growth of a market. Using this setup, we derive the relationship between the growth of a constant re-balanced portfolio as described above and the factor growth rates. For literature on factor based models in general see \cite{fama1993common} and \cite{jegadeesh1993returns} particularly for momentum factor related effects. The factor model for asset $j$ we use is:
\begin{equation*}\label{eq:gain}
R_{ji} = (r_{1i}) ^ {\beta^{1}_{j}} (r_{2i})^ {\beta^{2}_{j}} \quad j \in (1...m)
\end{equation*}
where $r_{1i}$ and $r_{2i}$ are returns attributed to factor 1 and factor 2 and $\beta$ denotes the factor exposures. 
The average growth rate of this portfolio is given by:
\begin{equation*}\label{eq:growth6}
GR_{avg} = \frac{1}{n} \, E \, \sum_{i=1}^{n} \log (\sum_{j=1}^{m} b_j R_j)  
\end{equation*}

\begin{equation*}\label{eq:growth7}
GR_{avg} = \frac{1}{n} \, E \, \sum_{i=1}^{n} \log (\sum_{j=1}^{m} b_j r_{1i} ^ {\beta^{1}_{j}} r_{2i}^ {\beta^{2}_{j}}  )  
\end{equation*}
By the AM-GM inequality and by the property that log is a monotone function:
\begin{equation*}\label{eq:growth8}
GR_{avg} \geq \frac{1}{n} \, E \, \sum_{i=1}^{n} \log ( m \prod_{j=1}^{m}  ( b_j r_{1i} ^ {\beta^{1}_{j}} r_{2i}^ {\beta^{2}_{j}} )^{\frac{1}{m}}  )  
\end{equation*}

\begin{equation*}\label{eq:growth9}
GR_{avg} \geq \log m + \frac{1}{mn} E \left[    \sum_{i=1}^{n}  \sum_{j=1}^{m}    \log          ( b_j r_{1i} ^ {\beta^{1}_{j}} r_{2i}^ {\beta^{2}_{j}} )                                             \right] 
\end{equation*}

\begin{equation*}\label{eq:growth10}
GR_{avg} \geq \log m + \frac{1}{mn} E \left[    \sum_{i=1}^{n}  \sum_{j=1}^{m}    \log b_j +   \beta^{1}_{j} \log r_{1i}  +  \beta^{2}_{j} \log r_{2i}                                                       \right] 
\end{equation*}

\begin{align*}\label{eq:growth11}
GR_{avg} &\geq \log m + \frac{1}{m} \sum_{j=1}^{m}    \log b_j +  \frac{1}{m}  \sum_{j=1}^{m}  \beta^{1}_{j} E \left[ \frac{1}{n} \sum_{i=1}^{n} \log (r_{1i})  \right]   \\  
&+ \frac{1}{m}  \sum_{j=1}^{m}  \beta^{2}_{j} E \left[ \frac{1}{n} \sum_{i=1}^{n} \log (r_{2i})  \right]
\end{align*}

\begin{equation*}\label{eq:growth12}
GR_{avg} \geq \log m + \frac{1}{m} \sum_{j=1}^{m}    \log b_j +  \frac{1}{m}  \sum_{j=1}^{m}  \beta^{1}_{j} G_1 +  \frac{1}{m}  \sum_{j=1}^{m}  \beta^{2}_{j} G_2
\end{equation*}

\begin{equation*}\label{eq:growth13}
GR_{avg} \geq  \beta_{1}^{AM} G_1 + \beta_{2}^{AM} G_2 - \log \frac{1}{ m \, b^{GM} }  
\end{equation*}
where AM stands for arithmetic mean and GM stands for geometric mean and $G_1$ and $G_2$ are buy and hold returns from factor 1 and factor 2 respectively.  
The main insight from the result is that the growth rate of a portfolio that re-balances regularly can do better than the underlying factor growth rates, depending on the average $\beta$ exposure of the assets underlying the portfolio and how diversified the portfolio is. The last term denotes a portfolio concentration cost term that vanishes for a $1/m$ diversified portfolio.

\subsection{Universality of Dirichlet factor portfolios}
Next we present a modification of Cover's\cite{Cover91} universal portfolio selection process to incorporate economic information. We focus on the alpha parameter of the dirichlet sampling on the unit simplex. Recall that alpha is a concentration parameter of the dirichlet distribution, in the sense that a diffuse alpha leads to portfolios with no prior knowledge. On the other extreme, an alpha where one component is very large and the others very small would tend to bias the portfolio towards that corresponding asset. We propose using dirichlet($\alpha=\beta$) distributions, where $\beta$ are factor loadings of assets, to create universal portfolios that perform better than the benchmark universal portfolio that \cite{Cover91} suggests. To show that the portfolio is universal, recall that Cover and Ordentlich\cite{cover1996universal} in result (24) in their paper show that the total wealth generated by the dirichlet $(1,1,1...1)$ distribution is given as:

\begin{equation*}\label{eq:dirichlet0}
W_{\mathbf{1} } = E_{F(x),b} \left[ \int_{b \in D_{\mathbf{1}}} S_n(b, X^n) \; d \mu(b) \right]
\end{equation*}
where 

\begin{equation*}\label{eq:dirichlet1}
S_n(b, X^n) = \prod_{i=1}^{n} b^t X_i
\end{equation*}

The method that we propose involves the same process, and samples repeatedly from a dirichlet distribution ($\alpha = \beta$) that is driven by factor exposures that each asset has to the underlying factor:

\begin{equation*}\label{eq:dirichlet2}
W_{\mathbf{\beta} } = E_{F(x),b} \left[ \int_{b \in D_{\mathbf{\beta}}} S_n(b, X^n) \; d \mu(b) \right]
\end{equation*}

Keeping in mind that Cover and Ordentlich\cite{cover1996universal} show that the dirichlet $(1,1...,1)$ process leads to a portfolio that is universal, we are interested in analyzing the properties of the ratio of these wealth processes. A ratio where the denominator is the naive dirichlet wealth with no information and the numerator is the new informed dirichlet wealth with knowledge of the factor $\beta$s describes the relative performance of these portfolios:

\begin{equation*}\label{eq:dirichlet3}
R_n =\frac{E_{F(x),b} \left[ \int_{b \in D_{\mathbf{\beta}}} S_n(b, X^n) \; d \mu(b) \right] }{ E_{F(x),b} \left[ \int_{b \in D_{\mathbf{1}}} S_n(b, X^n) \; d \mu(b) \right]}   
\end{equation*}

\begin{equation*}\label{eq:dirichlet4}
R_n =\frac{   \int_{b \in D_{\mathbf{\beta}}}  E_{F(x),b} \left[ S_n(b, X^n) \right] \; d \mu(b)  }{   \int_{b \in D_{\mathbf{1}}} E_{F(x),b} \left[ S_n(b, X^n) \right] \; d \mu(b)  }   
\end{equation*}

\begin{equation*}\label{eq:dirichlet5}
R_n =\frac{   \int_{b \in D_{\mathbf{\beta}}}  E_{F(x),b} \left[ 
\prod_{i=1}^{n} b^t X_i
 \right] \; d \mu(b)  }{   \int_{b \in D_{\mathbf{1}}} E_{F(x),b} \left[ 
\prod_{i=1}^{n} b^t X_i
 \right] \; d \mu(b)  }   
\end{equation*}

We proceed with the goal of explicitly showing that this ratio converges to a number larger then one, which would show that the wealth gains from the Dirchlet($\beta_1,\beta_2,\beta_3...\beta_m$) in expectation dominates the gains from the dirichlet(1,1,1...1) portfolio.
Recall that the one-factor model led to a relationship between the asset return and market as follows:

\begin{equation*}\label{eq:factormodel2}
\begin{bmatrix}
X_1\\
X_2\\
.\\
.\\
X_m\\
\end{bmatrix} = \begin{bmatrix}
\beta_{1} \; R_{m}\\
\beta_{2}\; R_{m}\\
.\\
.\\
\beta_{m}\; R_{m}\\
\end{bmatrix}  
\end{equation*}
This leads to the ratio being evaluated as:

\begin{equation*}\label{eq:dirichlet6}
R_n =\frac{   \int_{b \in D_{\mathbf{\beta}}}  E_{F(x),b} \left[ 
\prod_{i=1}^{n} R_{m}^{i} \prod_{i=1}^{n} b_{i}^{t} \beta
 \right] \; d \mu(b)  }
 {   \int_{b \in D_{\mathbf{1}}}  E_{F(x),b} \left[ 
\prod_{i=1}^{n} R_{m}^{i} \prod_{i=1}^{n} b_{i}^{t} \beta
 \right] \; d \mu(b)  }
\end{equation*}

\begin{equation*}\label{eq:dirichlet7}
R_n =\frac{   \int_{b \in D_{\mathbf{\beta}}}  E_{F(x),b} \left[ 
\prod_{i=1}^{n} R_{m}^{i}\right]   \prod_{i=1}^{n}  E_{F(x),b}  \left[  b_{i}^{t} \right] \beta
  \; d \mu(b)  }
 {   \int_{b \in D_{\mathbf{1}}}  E_{F(x),b} \left[ 
\prod_{i=1}^{n} R_{m}^{i}\right]   \prod_{i=1}^{n}  E_{F(x),b}  \left[  b_{i}^{t} \right] \beta
  \; d \mu(b)  }
\end{equation*}

\begin{equation*}\label{eq:dirichlet8}
R_n =\frac{   \int_{\Delta}  E_{F(x),b^{\beta}} \left[ 
\prod_{i=1}^{n} R_{m}^{i}\right]   \prod_{i=1}^{n}  E_{F(x),b^{\beta}}  \left[  b_{i}^{\beta} \right] ^{t} \beta
  \; d \mu(b)  }
 {   \int_{\Delta}  E_{F(x),b^{1}} \left[ 
\prod_{i=1}^{n} R_{m}^{i}\right]   \prod_{i=1}^{n}  E_{F(x),b^{1}}  \left[  b_{i}^{1} \right]^{t} \beta
  \; d \mu(b)  }
\end{equation*}

\begin{equation*}\label{eq:dirichlet9}
R_n =\frac{   \int_{\Delta}         \left[  m^n (\sum_{j=1}^{m} \beta_{j}^{2})^n    \right]  
  \; d \mu(b)  }
 {   \int_{\Delta}         \left[   (\sum_{j=1}^{m} \beta_{j} ) ^{2n} \right]
  \; d \mu(b)  }
\end{equation*}

which is a ratio of two integrals over the unit (non-negative) simplex where the numerator always dominates the denominator: 
\begin{equation}\label{eq:centralresult}
F_n = \frac{m^n (\sum_{j=1}^{m} \beta_{j}^{2})^n  }{ (\sum_{j=1}^{m} \beta_{j} ) ^{2n}} \geq 1
\end{equation}
By Cauchy-Schwartz $F_n$ is greater than 1. Consequently, $R_{n}$ is lower bounded by 1 and we have shown that the new factor tilted portfolio proposed is also universal. 

\section{Empirical Results: Dirichlet factor portfolios}
We highlight the practical consequences of our results by showing four use cases across two separate equity market settings, the Nasdaq 100 and the FTSE 100. For both asset universes, we show that it is possible to construct factor dirichlet portfolios for factors that are already known in the literature to have an impact on the cross-section of returns for that asset class.

One such factor is size, where there is abundant literature to show that small companies outperform their larger counterparts\cite{fama1993common}. As a simple proxy for size, we use the traded price at any point in time as an indication of size. While this can be critiqued, it serves as a simple calculation keeping our information set requirement within the pricing domain. Note that to get to a set of weights that be used for the dirichlet sampling, a few cross-sectional normalizations are also needed. As a starting example, we use the transformations shown in Algorithm 1 to obtain an alpha value for our dirichlet simulation at every step.

\begin{algorithm}
\label{algo}
\caption{Calculating portfolio weights for SizeFactor based dirichlet portfolio}\label{alg:cap}
\begin{algorithmic}
\While{$t \neq T$}
\State $ \alpha_{jt} \gets 1/p_{jt} $
\State $ \alpha_{jt} \gets \frac{\alpha_{jt}}{min(\alpha_{jt})} $
\Require Simulate N-Managers dirichlet portfolios from Dirichlet($\alpha = \alpha_{t}$)
\Require Calculate Universal Portfolios from factor biased managers and update the wealth process
\State $ w_{jt} \gets \frac{\int(S b db)}{\int(S db)}   $
\State $ t \gets t+1$
\EndWhile
\end{algorithmic}
\end{algorithm}

The result of this algorithm is a portfolio that shows favorable characteristics. We show the results for a few such factor portfolios in Figure 1 for Nasdaq 100 constituents from August 2007 to June 2023. Results are repeated for FTSE 100 constituents in Figure 2.

\subsection{Nasdaq 100 Constituents}
\label{ex3}
In this example we used a selection of 71 assets (listed cash equity) from the US stock market that are a subset of the Nasdaq 100 index constituents. Tickers that were part of the index as of June 6th, 2023 and had availability of daily price time-series data for at least 4,000 trading days were chosen as the asset universe. The tickers were retrieved from the Nasdaq web page \cite{nasdaqWebsite} 
on the same date, 101 tickers were filtered using the above criteria to yield 71 tickers with at least the minimum threshold of price history. Daily price data was downloaded from EOD HistoricalData's (a commercial service) end of day API \cite{eodWebsite}
. We down-sampled the daily price data to a weekly time-series and used the weekly median price as the entry and exit price for the trading simulation.\\

Table \ref{factors} shows the computational functional form behind each of the factors shown in the charts. As we have access only to price data for this study, we limit the experiment to only price-based factors. As an example, the momentum factor is calculated as a function of the cumulative return of the asset (r). In this case, we cross-sectionally z-score(Z) the cumulative return (r) to date, winsorize(w) the z-score between -6 and 6, exponentiate it and then truncate the result to be always greater than 1. These back of the envelope functions are meant to be easy to calculate and at the same time intend to mimic the factor target. They also need to adhere to the allowed values for a dirichlet distribution $\alpha$ parameter input.

\begin{table}[!htbp] 
\small
\centering 
  \caption{Factor based Dirichlet Alpha} 
  \label{factors} 
\begin{tabular}{@{\extracolsep{5pt}} ccc} 
\\[-1.8ex]\hline 
\hline \\[-1.8ex] 
 & FactorType & Dirichlet Alpha \\ 
\hline \\[-1.8ex] 
1 & Size & $\left [ \frac{\frac{1}{p}}{min(\frac{1}{p})}) \right ] $  \\ 
2 & Mom ($\mu$) & $\max \left\{e^{w(Z(r), 6)}, 1 \right\}  $  \\ 
3 & Sharpe ($\nu$) & $\max \left\{w(Z(S(r,10)), 6), 1 \right\}$  \\ 
4 & Compound & $\sqrt{\mu * \nu}$  \\

\hline \\[-1.8ex]
\end{tabular} 
\label{dd1table}
\end{table} 

$S$ in the sharpe factor stands for the rolling sharpe ratio, which is computed using a 10 week span exponentially weighted moving average of the returns divided by the 10 week span rolling standard deviation of the returns. In each case the most recent available metric is used.
We also illustrate a compound factor that is the geometric mean of the momentum and sharpe factor, which opens up the possibility of constructing functional forms that depend on a variety of factors.

\begin{figure}[ht]
\centering
\includegraphics[scale=0.40]{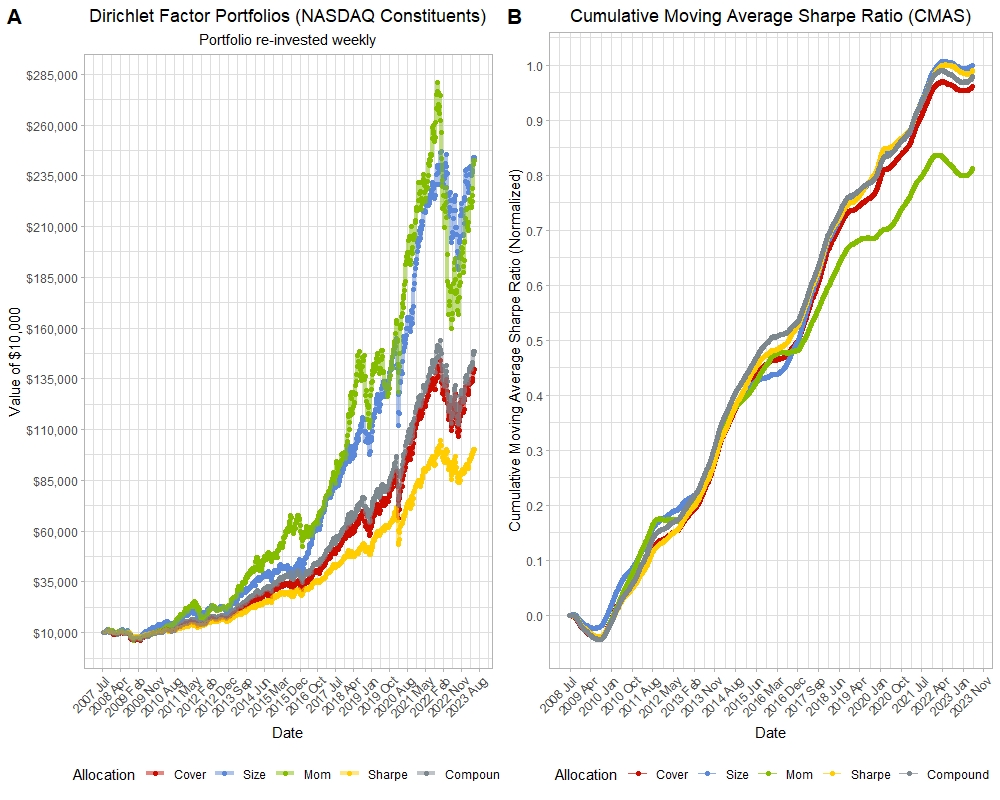}
\caption{Backtest Results: Nasdaq 100 constituents with various dirichlet factors }
\label{nasdaq}
\end{figure}

The results are shown in Figure \ref{nasdaq}. Note that this back-test has an inherent look-ahead bias in that it uses assets that survive the period between August 2007 and June 2023. Hence the asset universe comprises of higher quality assets than the Nasdaq at that point in time. Since we are comparing only the relative performance of the class of naive-dirichlet portfolios to the factor biased dirichlet portfolios, this chart serves its purpose as a relative comparison and is not indicative of absolute performance. The back-test (Chart Figure 1A) shows out-performance both for the momentum and size dirichlet portfolios in terms of the growth rate of the portfolio. The cumulative moving average sharpe ratio chart (CMAS - Figure 1B) shows that all factor return streams dominate Cover's portfolio on a sharpe ratio basis, except the momentum factor. A smoothing window of 50 days was used to calculate sharpe ratios for each curve and the result was cumulated and normalized to have a maximum value of 1.

Figure 2 shows the same effects for the constituents of the FTSE, a Europe based index.

\subsection{FTSE 100 Constituents}
Now we use a selection of 87 cash equity assets from the stocks that are listed on the London Stock Exchange (LSE) and are part of the FTSE 100 as of June 22nd 2023. The criteria for selection were the same as in the previous example with Nasdaq 100 stocks. Tickers that were part of the FTSE index as of June 22nd, 2023 and had availability of daily price time-series data for 4,000 trading days as of July 6th 2023 were chosen as the asset universe. In the FTSE dataset, the size effects are quite large. To control for this effect and not to let it influence our results too much, we remove assets that had a minimum price of less than \$1 through the time period considered. Only one such asset was removed from the dataset in this case.
The tickers were retrieved from the LSE web page \cite{ftseWebsite}. Trading conditions and factors are the same as what is described in the section on Nasdaq constituents. Similar outsize performance is observed for factor dirichlet portfolios, with the size based dirichlet portfolio dominating other allocations (Figure 2A). CMAS curves show regular out-performance in moving sharpe ratios for dirichlet portfolios as shown in Figure 2B.

\begin{figure}[ht]
\centering
\includegraphics[scale=0.40]{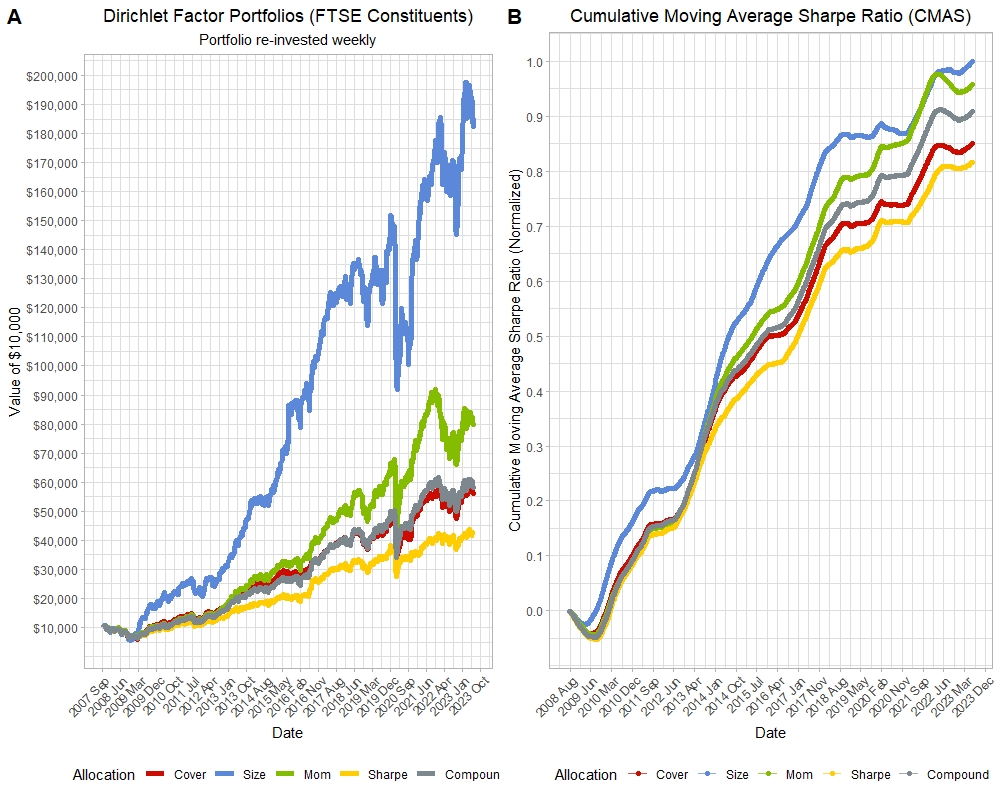}
\caption{Backtest Results: FTSE 100 constituents with various dirichlet factors }
\label{ftse}
\end{figure}

\section{Conclusion}
In this work, we have proposed a generative process to create robust, high sharpe, diversified, universal portfolios that outperform the naive un-informed dirichlet universal portfolio allocation. An analytical proof has been shown that the portfolios are indeed universal and they dominate portfolios that arise out of a uniform dirichlet sampling. Particular examples of this technique have been demonstrated on real market data from the U.S. and European equities market.  We believe that this technique can be used by small and large investment managers, robo-advisors and algorithmic trading outfits as a target portfolio for incremental alpha generation. Further work could involve online learning of the parameter for dirichlet inputs based on recent performance of factors in the market, and being able to blend them as shown in the composite portfolio. Another line of work could include blending price and non-price factors to create mixture portfolio processes.

\bibliographystyle{plainnat}
\bibliography{UDFP}

\ignore{

\section*{Appendix}
Lemma to show (\ref{eq:centralresult}) \label{lemma:cs} using Cauchy-Schwartz Inequality:

\begin{equation*}\label{eq:cs1}
\begin{Vmatrix}
u.v
\end{Vmatrix}^2 \leq \begin{Vmatrix}
u.u
\end{Vmatrix}
\begin{Vmatrix}
v.v
\end{Vmatrix}
\end{equation*}

Setting $u$ as $\mathbf{1}$ and $v$ as $\beta$, both in $\mathbf{R^m}$
\begin{equation*}\label{eq:cs2}
\begin{Vmatrix}
\mathbf{1}.\beta 
\end{Vmatrix}^2 \leq \begin{Vmatrix}
\mathbf{1}.\mathbf{1}
\end{Vmatrix}
\begin{Vmatrix}
\beta .\beta
\end{Vmatrix}
\end{equation*}

\begin{equation*}\label{eq:cs3}
\begin{Vmatrix}
\mathbf{1}.\beta 
\end{Vmatrix}^2 \leq m \sum_{i=1}^{m} \beta^2
\end{equation*}

\begin{equation*}\label{eq:cs4}
(\sum_{i=1}^{m} \beta_j)^2 \leq m \sum_{i=1}^{m} \beta_j^2
\end{equation*}
\qedsymbol
}

\end{document}